\documentclass[usletter,
               keeplastbox,   
               ]{jacow}
\usepackage{pdfpages,multirow,ragged2e} %
\makeatletter%
	\ifboolexpr{bool{xetex}}
	 {\renewcommand{\Gin@extensions}{.pdf,%
	                    .png,.jpg,.bmp,.pict,.tif,.psd,.mac,.sga,.tga,.gif,%
	                    .eps,.ps,%
	                    }}{}
\makeatother

\ifboolexpr{bool{xetex} or bool{luatex}} 
 {}                                      
 {\usepackage[utf8]{inputenc}}           

\usepackage[USenglish]{babel}

\ifboolexpr{bool{jacowbiblatex}}%
 {%
  \addbibresource{jacow-test.bib}
  \addbibresource{biblatex-examples.bib}
 }{}
\graphicspath{{./figures/}}
\begin{document}

\title{Space charge dominated momentum spread and compensation strategies in the post-linac section of Proton Improvement Plan-II at Fermilab\thanks{This manuscript has been authored by Fermi Research Alliance, LLC under Contract No. DE-AC02-07CH11359 with the US Department of Energy, Office of Science, Office of High Energy Physics.
}}

\author{A. Pathak\thanks{abhishek@fnal.gov}, O. Napoly\thanks{napoly@fnal.gov}, J.-F. Ostiguy\\ Fermi  National Accelerator Laboratory, Batavia, USA  }
	
\maketitle
\begin{abstract}
The upcoming Proton Improvement Plan-II (PIP-II), designated for enhancements to the Fermilab accelerator complex, features a new 800 MeV superconducting linac and a Beam Transfer Line (BTL) to transport the beam to the existing Booster synchrotron. To mitigate the space charge tune shift associated with a high intensity accumulated beam, the low emittance linac beam is used to paint the ring phase space both transversely and longitudinally. 
To prevent losses caused by particles injected outside the rf separatrix while painting longitudinal phase space, the momentum spread of the incoming beam should not exceed $2.1\times 10^{-4}$. 
Detailed simulations showed that due to space charge, the rms momentum spread increases to $4\times 10^{-4}$ while it is transported in the BTL --about twice the allowable limit. 
In this paper, we outline a mitigation strategy involving a debuncher cavity. We discuss location, operating frequency, and gap voltage under both nominal and perturbed beam conditions, specifically accounting for momentum jitter. The impact of  cavity misalignments is also assessed. The paper concludes by recommending an optimized configuration.
\end{abstract}
\section{INTRODUCTION}
The Fermilab Proton Improvement Plan-II (PIP-II)\cite{cdr} represents a pivotal upgrade to the Fermilab accelerator complex, specifically designed to enhance the capabilities of the Deep Underground Neutrino Experiment (DUNE) at the Long-Baseline Neutrino Facility (LBNF). The upgrade involves construction of new superconducting linac\cite{PIP-II} which will accelerate a 2 mA, H$^-$ beam to an energy of 800 MeV for accumulation by charge exchange in the existing Booster synchrotron. The beam is injected via a Beam Transfer Line (BTL)  as depicted in Fig. 1(a). The transfer line is based on 90 degree FODO cells. Because the new injection point has to be situated on the side of the ring opposite to the previous one, it comprises two large circular arcs separated by an eight cell dispersion free straight section. The circular arcs are used to reverse the beam direction to allow for injection compatible with the established ring circulation direction.  

\begin{figure}
\includegraphics[width=80mm, draft=false]{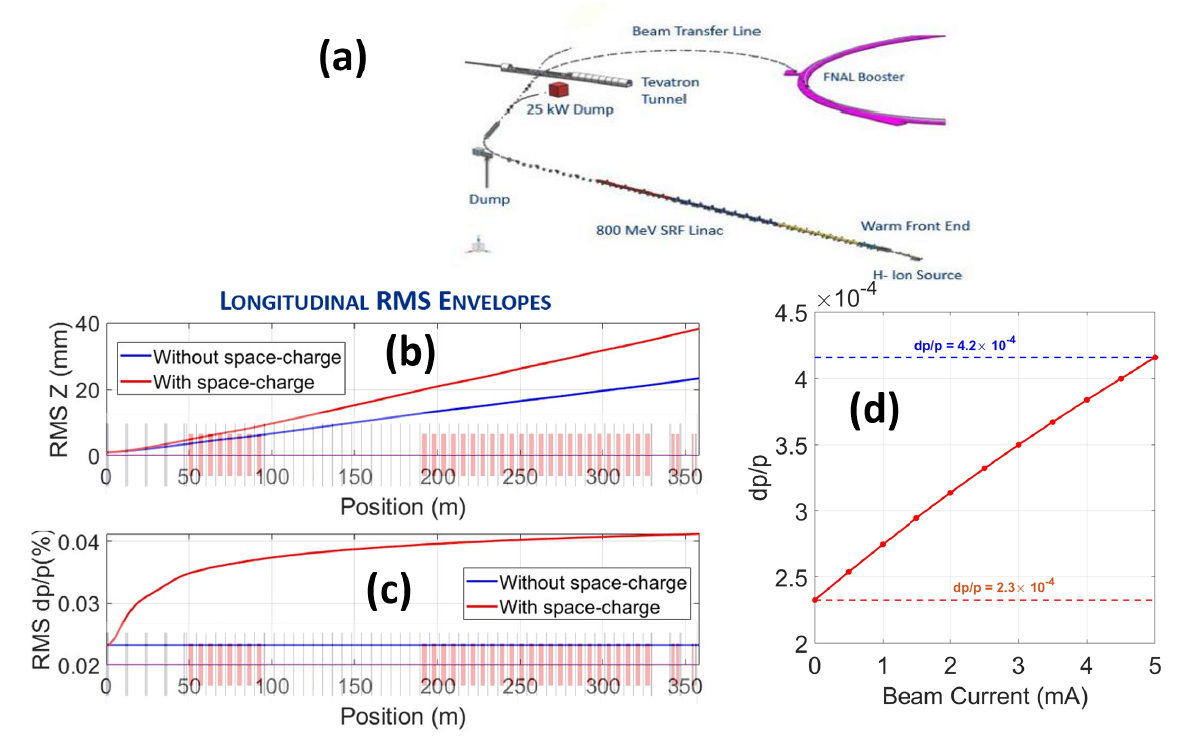}
\caption{(a) Schematic of the SC linac and beam transport to the booster via BTL, (b) Variation of longitudinal bunch size along BTL, (c) Increase in $dp/p$ along BTL, and (d) Variation of $dp/p$ with beam current.}
\label{Fig1}
\end{figure}

Upon exiting the SC linac, the 800 MeV beam is not fully relativistic ($\beta = 0.84$) and still influenced by space charge, with a generalized perveance of \(8.42 \times 10^{-12}\). Simulations indicate that in the  absence of additional measures, space charge causes momentum spread to double, from \(2.1 \times 10^{-4}\) to \(4.2 \times 10^{-4}\) (Fig. \ref{Fig1}(c)). 


\section{Debuncher Cavity}
As shown in Fig. 1(b), upon leaving the SC linac, the momentum spread increases due to space charge forces. As the beam spreads out, the charge density progressively decreases, and the momentum spread stabilizes. An rf cavity operating at phase $\phi_s=0$  (referred to as a debuncher cavity in this context for its role in reducing $dp/p$ growth), can be used to counteract space charge defocusing. 
Treating the cavity as a simple rf gap, the focusing strength is obtained as follows\\
\begin{eqnarray}
    \frac{\Delta p}{p} & = & \frac{q{\cal E}_0 T L}{\beta^2 E} \sin \left[ \omega \frac{\Delta z}{\beta c} + \phi_s \right]\\
    & \simeq & \frac{q\omega {\cal E}_0 T L }{\beta^3Ec} \Delta z = \frac{q\omega V} {\beta^3Ec}  \Delta z
\end{eqnarray}
where $\Delta z$ is the longitudinal position with respect to the reference particle, $\omega$ is the rf frequency $T$ is the transit time factor, ${\cal E}_0$ is the electric field amplitude, $L$ is the cavity effective length, $E$ is to total energy, $q$ is particle charge and $V$ is the gap voltage. The focusing strength is proportional to the product of the voltage and the frequency. The focusing range is limited to $|\Delta z| < \lambda/4 $ where $\lambda = \frac{2\pi\beta c}{\omega}$ is the rf wavelength.  
\subsection{Placement}
As the bunch length increases away from the linac, at given voltage and frequency rf focusing becomes more effective with distance.  This favors positioning the cavity closer to the downstream end of the beamline. However, once the length $L_z$ of the drifting bunch reaches $\sim \lambda/4$,  focusing becomes nonlinear and less effective and is eventually lost when $L_z \simeq \lambda/2$.  

\begin{figure}
\includegraphics[width=80mm, draft=false]{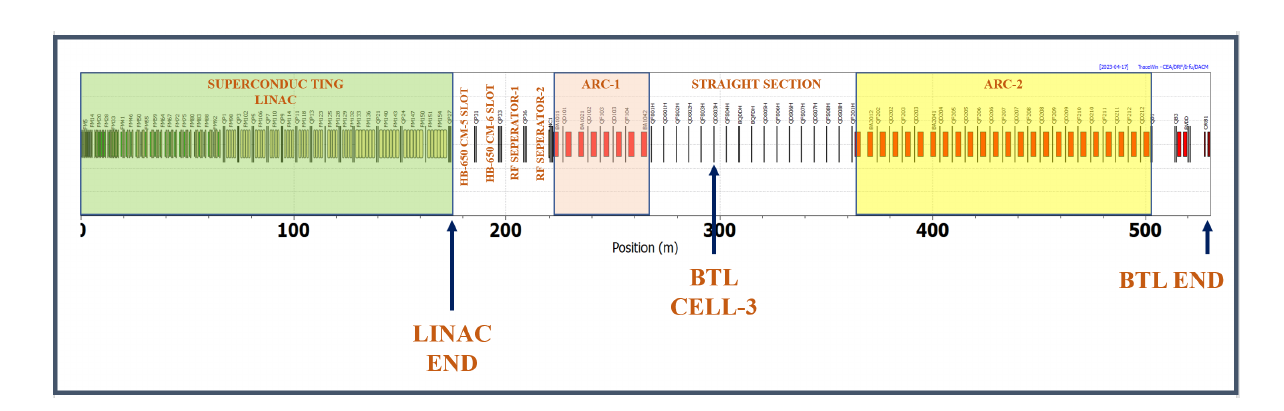}
\caption{Schematic diagram illustrating the superconducting linear accelerator and Beam Transfer Line from the SC linac to the Booster.}
\label{Fig2}
\end{figure}

The need for sufficient available longitudinal space restricts the possible location for a cavity to the straight section. Many of the cells within this section are claimed for by diagnostics, transverse collimation and other components. Taking in to account factors such as accessibility and space constraints, only cell-3 was considered suitable to house a cavity, as shown in Fig. \ref{Fig2}. Consequently, this specific location was adopted for further investigations.
\subsection{Frequency}
\begin{figure}
\centering
\includegraphics[width=85mm, draft=false]{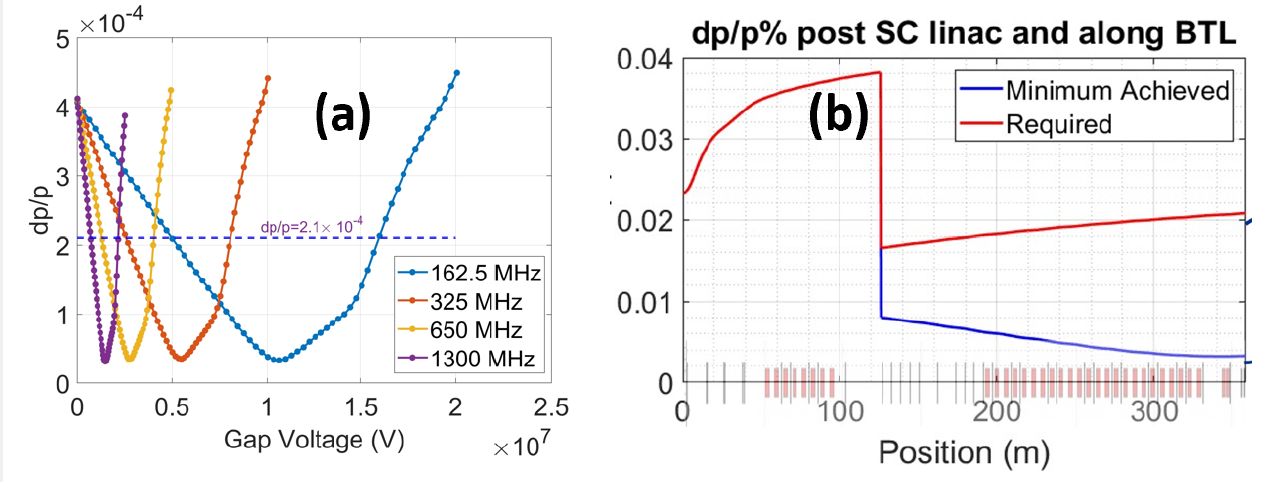}
\caption{(a)Variation in momentum spread at the exit of the BTL, as a function of the debuncher gap voltage operating at different frequencies, (b) Variation of momentum spread along the BTL after momentum spread compensation using the 650 MHz cavity in Cell-3 of BTL.}
\label{Fig3}
\end{figure}

With the cavity location established, the next step is the selection of an operating frequency. Simulations were carried out to quantify the gap voltage necessary to achieve the specified momentum spread required for booster injection ($2.1\times 10^{-4}$). The analysis considered cavities operating at frequencies of 162.5, 325, 650, and 1300 MHz. For each of these frequencies, the momentum spread at the exit of the BTL, as a function of the gap voltage, is illustrated in Fig. \ref{Fig3}(a).\\
As depicted in Fig. 3(a), higher operating frequency of the cavity leads to more effective focusing, thereby requiring a lower gap voltage to achieve a given reduction in momentum spread growth. Within the examined frequency range, operating frequencies of 1300 MHz and 650 MHz yield the most promising results, necessitating lower gap voltages of 0.7 MV and 1.3 MV, respectively. These requirements are significantly lower than the 2.7 MV and 5 MV needed for cavities operating at 325 MHz and 162.5 MHz.
Taking into account factors such as cavity size and the higher gap voltage requirements, the 162.5 MHz and 325 MHz cavities were excluded from further consideration. A comparative analysis of the advantages and disadvantages between the 1300 MHz and 650 MHz cavities revealed that while the 1300 MHz cavity achieves the desired $dp/p$ with only 54\% of the voltage needed at 650 MHz, it would require a large cumbersome and expensive rf power source. Resorting to superconducting technology would somewhat alleviate this; however, there is no cooling infrastructure readily available in the BTL tunnel. In contrast the warm 650 MHz option allows for more compact and economical rf power with reasonable cavity dimensions, simplifying operational requirements. Taking these factors into account, and with a modest compromise on the gap voltage, the decision was made to proceed with a room-temperature cavity operating at 650 MHz. The evolution of the rms $dp/p$ along the BTL with the 650 MHz cavity operating at gap voltages of 1.3 MV and 2.9 MV  respectively corresponding to the required and minimum achievable momentum spreads is presented in Fig. \ref{Fig3}(b). Fig. \ref{Fig3} clearly showcases the effectiveness of the debuncher cavity in compensating momentum spread under ideal conditions. 
\section{Tolerances}
Reliability and tolerance to element misalignments and beam parameter deviations are important factors to consider. In addition to cavity misalignments, the beam emerging from the SC linac is subject to intensity fluctuations, emittance variations and momentum jitter. 
We performed detailed simulations to investigate the impact of such factors on the longitudinal phase space at the upstream extremity of the BTL.
\subsection{Beam Parameters}
\begin{figure}
\centering
\includegraphics[width=70mm,draft=false]{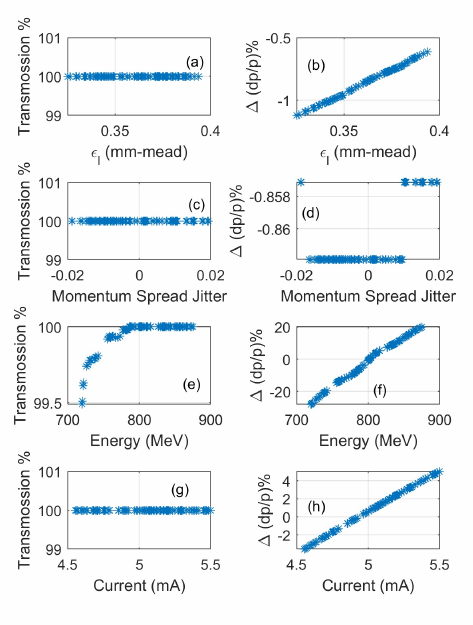}
\caption{Variability of beam transmission and momentum spread. (a, c, e, g) display the transmission percentage against variations in longitudinal emittance, momentum jitter, energy, and current. (b, d, f, h) illustrate the impact of these parameters on momentum spread at the exit of the BTL, indicating parameter sensitivity.}
\label{Fig4}
\end{figure}

Assuming a debuncher cavity  optimized for nominal operational conditions, 10\% variations in beam current, emittance, energy, and momentum were introduced and variations in momentum spread and beam transmission at the exit of the BTL were observed. The findings are presented in Fig.\ref{Fig4}.
As the linac energy is reduced from 800 MeV to 720 MeV, a marginal influence on beam transmission is observed, with losses not exceeding 0.5\%. These losses may be attributed to transverse mismatch induced by alterations in beam energy. No discernible effect on transmission were observed in relation to other beam parameters. The momentum spread at the BTL exit is linearly dependent on  beam energy,  current and longitudinal emittance, characterized by slopes of $10^{-5}$ (1/MeV), $1.9 \times 10^{-5}$  (1/mA), and $1.6 \times 10^{-5}$ (1/mm-mrad), respectively. Notably, when a linac central momentum jitter $ \left|\frac{dp}{p}\right| \leq 2 \times 10^{-4}$ is introduced, no corresponding meaningful variation at the output of the BTL is observed, indicating that the debuncher cavity is effective at suppressing jitter as well as momentum spread. 
\subsection{Misaligments, Phase and Amplitude Errors}
\begin{figure}
\centering
\includegraphics[width=67mm, draft=false]{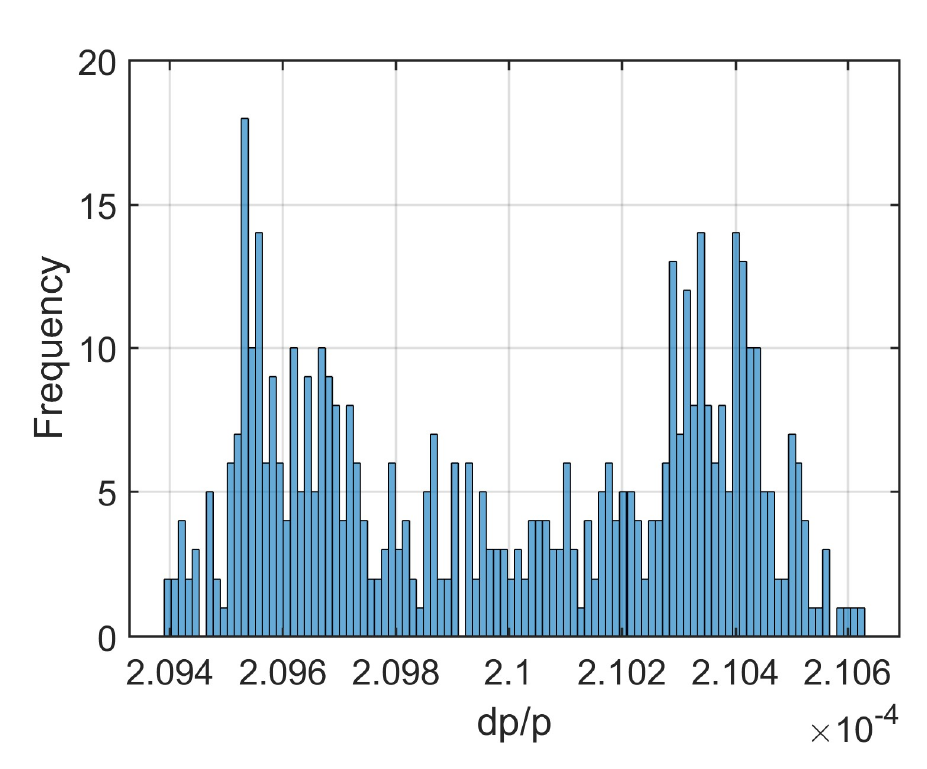}
\caption{Impact of debuncher cavity parameter errors on  $\frac{dp}{p}$ variation at the upstream end of the BTL. This figure illustrates the effect of simultaneously introducing uniform random misaligment errors $\Delta X,\Delta Y, \Delta Z = \pm 4$ mm offsets, and a 5\% error in both field amplitude and phase, across 500 simulation runs.}
\label{Fig5}
\end{figure}

We investigated the effects of debuncher misalignments as well as those of phase and field amplitude jitter. Uniformly distributed random misaligments $\Delta X, \Delta Y, \Delta Z = \pm 4$  mm were introduced, together with a 5\% in both field amplitude and phase. All errors were introduced simultaneously and randomly over an ensemble of 500 runs. Fig. \ref{Fig5} shows variations in beam transmission and $\frac{dp}{p}$ attributable to the specified errors in cavity parameters.
As depicted in Fig. \ref{Fig5}, in the presence of cavity misalignments within \(\pm 4 \, \text{mm}\),  field amplitude and phase errors within $\pm 5$\% and $\pm 5$ deg, using a nominal momentum spread value of \(2.1\times 10^{-4}\) we observe a very small variation of 0.3\% at the exit of the BTL. The beam transmission remains unaffected. 
%
\section{Conclusion}
Our investigation shows that a debuncher cavity is effective at reducing  momentum spread from $4.2 \times 10^{-4}$ to within the Booster's allowable limit of $2.1 \times 10^{-4}$. 
Detailed simulations indicate that beam parameter variations and cavity misalignments have minimal impact on the emerging phase-space distribution. Specifically, beam parameter variations did not result in meaningful transmission losses (not exceeding 0.5\%). The momentum spread remained within a narrow variance of 0.3\% around the nominal $2.1 \times 10^{-4}$, despite cavity misalignments up to $\pm 4\, \text{mm}$ combined to field amplitude and phase errors of $\pm 5$\% and $\pm 5$ deg.  Importantly, a cavity is effective at reducing both the bunch momentum spread and variations in average momentum due to momentum jitter originating from the SC linac; both contribute to the total effective momentum spread relevant to phase space painting.

---

%
\ifboolexpr{bool{jacowbiblatex}}%
	{\printbibliography}%
	{%
	

}
\end{document}